\documentclass{article}
\usepackage[T1]{fontenc}
\usepackage[latin1]{inputenc}
\usepackage{amsmath}

\makeatletter

\providecommand{\LyX}{L\kern-.1667em\lower.25em\hbox{Y}\kern-.125emX\@}

\usepackage{amssymb}
\usepackage{amsfonts}
\makeatother

\begin{document}

\title{Symmetric products, permutation orbifolds and discrete torsion}

\author{P. Bantay\\
Inst. for Theor. Phys.\\
Rolland Eötvös Univ.}

\maketitle
\begin{abstract}
Symmetric product orbifolds, i.e. permutation orbifolds of the full symmetric
group \( S_{n} \) are considered by applying the general techniques of permutation
orbifolds. Generating functions for various quantities, e.g. the torus partition
functions and the Klein-bottle amplitudes are presented, as well as a simple
expression for the discrete torsion coefficients.
\end{abstract}
The importance of symmetric product orbifolds had been recognized long ago,
e.g. for second quantized strings \cite{DMVV},\cite{D} and matrix string theory
\cite{DVV}. From a geometric point of view, they amount to passing from a sigma
model describing string propagation on some target manifold \( X \) to the
sigma model for the Hilbert-scheme of \( X \). As argued in \cite{D1}, the
corresponding CFT should describe quantum string theory on \( X \). Many results
about symmetric products have been worked out by essentially Hamiltonian techniques,
e.g. the structure of the state space, the elliptic genus and its automorphic
properties, etc. These results constitute a promising approach to a consistent
second quantized string field theory.

The aim of the present paper is to present the theory of symmetric products
from the general point of view of permutation orbifolds \cite{KS},\cite{BHS},\cite{PO}.
After all, symmetric products are nothing but permutation orbifolds for the
full symmetric groups \( S_{n} \). As we'll show, this allows to obtain quite
explicit expressions for practically all quantities of interest, e.g. for the
partition function on arbitrary surfaces. Most of the results are simple corrolaries
of a general combinatorial identity, Eq.(\ref{comb}). We also discuss the issue
of discrete torsion \cite{D}, and give a simple closed expressions for the
discrete torsion coefficients appearing in the torus partition function and
the Klein-bottle amplitude. It should be stressed that our approach answers
questions that seem hard to attack with the essentially Hamiltonian methods
used in \cite{DMVV} and \cite{D}.

Starting from a CFT \( \mathcal{C} \), one may take the product of \( n \)
identical copies of \( \mathcal{C} \), resulting in the tensor power \( \mathcal{C}^{\otimes n} \).
As any permutation of these identical copies is a symmetry of \( \mathcal{C}^{\otimes n} \),
one may orbifoldize the latter by the full symmetric group \( S_{n} \), resulting
in the permutation orbifold \( \mathcal{C}\wr S_{n} \) which is our object
of interest. The theory of permutation orbifolds allows in principle the computation
of all interesting quantities of \( \mathcal{C}\wr S_{n} \) from the knowledge
of \( \mathcal{C} \), in particular the characters of the primary fields, the
fusion rule coefficients, the modular matrices, etc. \cite{PO}. Instead of
working with one symmetric group at a time, it turns out to be more convenient
to deal with all symmetric groups at once, expressing the results in terms of
generating functions \cite{DMVV},\cite{D1}. 

Most of the results of this paper follow from the general combinatorial identity
\begin{equation}
\label{comb}
\sum _{n=0}^{\infty }\frac{1}{n!}\sum _{\phi :G\rightarrow S_{n}}\prod _{\xi \in \mathcal{O}(\phi )}\mathcal{Z}(G_{\xi })=\exp \left( \sum _{H<G}\frac{\mathcal{Z}(H)}{\left[ G:H\right] }\right) 
\end{equation}
 where \( G \) is any finitely generated group, while \( \mathcal{Z} \) is
a function on the set of finite index subgroups of \( G \) that takes its values
in a commutative ring and is constant on conjugacy classes of subgroups. The
second summation on the lhs. runs over the homomorphisms \( \phi :G\rightarrow S_{n} \)
from \( G \) into the symmetric group \( S_{n} \). For a given \( \phi  \),
we denote by \( \mathcal{O}(\phi ) \) the orbits of the image \( \phi (G) \)
on the set \( \left\{ 1,\ldots ,n\right\}  \), and \( G_{\xi }=\left\{ x\in G\, |\, \phi (x)\xi ^{*}=\xi ^{*}\right\}  \)
is the stabilizer subgroup of any point \( \xi ^{*}\in \xi  \)  of the orbit
\( \xi  \) - note that the lhs. of Eq.(\ref{comb}) is well defined, since
the stabilizers of points on the same orbit are conjugate subgroups. Finally,
\( \left[ G:H\right]  \) denotes the index of the subgroup \( H<G \), and
the \( n=0 \) term on the lhs. of Eq.(\ref{comb}) equals \( 1 \) by convention.

Although some special instances of Eq.(\ref{comb}) appeared previously in the
mathematical literature, e.g. 
\begin{equation}
\label{stanley}
\sum _{n=0}^{\infty }\frac{p^{n}}{n!}\sum _{\phi :G\rightarrow S_{n}}1=\exp \left( \sum _{H<G}\frac{p^{\left[ G:H\right] }}{\left[ G:H\right] }\right) 
\end{equation}
 for the generating function of the number of homomorphisms from \( G \) to
\( S_{n} \) \cite{Stanley} - this corresponds to \( \mathcal{Z}(H)=p^{\left[ G:H\right] } \),
where \( p \) denotes the formal variable of the power series ring \( \mathbb {C}\left\{ p\right\}  \)
-, to the best of our knowledge Eq.(\ref{comb}) has not appeared in print before.
If we take \( G \) to be the fundamental group \( \pi _{1}X \) of a manifold
\( X \), then the sum on the lhs. of Eq.(\ref{comb}) may be interpreted as
a sum over all unramified coverings of \( X \) suitably weighted, while the
exponent on the rhs. is a sum over connected coverings only. This last remark
explains the relevance of Eq.(\ref{comb}) to our problem. 

The CFT \( \mathcal{C} \) assigns a number \( Z(\mathcal{S}) \) - the partition
function - to each surface \( \mathcal{S} \). By the uniformization theorem
\cite{Farkas}, any surface \( \mathcal{S} \) may be obtained as the quotient
of its universal cover \( \hat{\mathcal{S}} \) by a suitable group \( G_{\mathcal{S}} \)
of automorphisms of \( \hat{\mathcal{S}} \), the uniformizing group, which
is determined up to conjugacy, and is isomorphic to the fundamental group \( \pi _{1}\mathcal{S} \)
of the surface. Let's define the values of \( \mathcal{Z} \) through the requirement
\[
\mathcal{Z}\left( G_{\mathcal{S}}\right) =Z\left( \mathcal{S}\right) \]
 Any finite index subgroup of \( G_{\mathcal{S}} \) uniformizes another surface,
which is a finite sheeted covering of \( \mathcal{S} \), so the above assignment
is meaningful. By the results of \cite{HG}, the quantity 
\begin{equation}
\label{partfun1}
Z_{n}\left( \mathcal{S}\right) =\frac{1}{n!}\sum _{\phi :G_{\mathcal{S}}\rightarrow S_{n}}\prod _{\xi \in \mathcal{O}(\phi )}\mathcal{Z}\left( G_{\mathcal{S},\xi }\right) 
\end{equation}
 is then nothing but the partition function of the surface \( \mathcal{S} \)
in the permutation orbifold \( \mathcal{C}\wr S_{n} \). Upon introducing the
notation 
\begin{equation}
\label{heck}
Z^{(n)}(\mathcal{S})=\frac{1}{n}\sum _{\left[ G_{\mathcal{S}}:H\right] =n}\mathcal{Z}(H)
\end{equation}
where the summation runs over the subgroups \( H<G_{\mathcal{S}} \) of index
\( n \), Eq.(\ref{comb}) gives 
\begin{equation}
\label{partgen}
\sum _{n=0}^{\infty }p^{n}Z_{n}(\mathcal{S})=\exp \left( \sum _{n=1}^{\infty }p^{n}Z^{(n)}\left( \mathcal{S}\right) \right) 
\end{equation}
for the generating function of the \( Z_{n} \)-s, where \( p \) denotes a
formal variable. The great advantage of this last formula is that the sum on
the rhs. of Eq.(\ref{heck}) is much more accessible to actual computations
than the one in Eq.(\ref{partfun1}). 

Let's note the following consequence of Eq.(\ref{partgen}) 
\begin{equation}
\label{partfun2}
Z_{n}(\mathcal{S})=\frac{1}{n!}\sum _{x\in S_{n}}\prod _{\xi \in \mathcal{O}(x)}Z^{(\left| \xi \right| )}(\mathcal{S})
\end{equation}
where \( \left| \xi \right|  \) denotes the length of the cycle \( \xi \in \mathcal{O}(x) \).
This should be contrasted with the expression Eq.(\ref{partfun1}) for the same
quantity, which involves a summation over all homomorphisms from \( G_{\mathcal{S}} \)
to \( S_{n} \), which is clearly much more involved then just summing over
elements \( x\in S_{n} \). 

The above results answer in complete generality the question of how to compute
partition functions in the permutation orbifolds \( \mathcal{C}\wr S_{n} \).
To get more familiar with them, let's consider the simplest case, namely the
torus partition function. In this case \( \mathcal{S} \) is a torus with modular
parameter \( \tau  \), and \( G_{\mathcal{S}} \) is isomorphic to \( \mathbb {Z}\oplus \mathbb {Z} \),
the fundamental group of the torus. The finite index subgroups \( H<\mathbb {Z}\oplus \mathbb {Z} \)
are characterized uniquely by a matrix in Hermite normal form, i.e. by a triple
of integers \( \left( \lambda ,\mu ,\kappa \right)  \), with \( \lambda \mu  \)
equal to the index of \( H \) and \( 0\leq \kappa <\lambda  \). The surface
uniformized by the subgroup \( H \) is again a torus, with modular parameter
equal to 
\[
\frac{\mu \tau +\kappa }{\lambda }\]
Taking this into account, and denoting \( Z(\mathcal{S}) \) by \( Z(\tau ) \)
as usual, we have 
\begin{equation}
\label{thecke}
Z^{(n)}\left( \mathcal{S}\right) =\frac{1}{n}\sum _{\lambda |n}\sum _{0\leq \kappa <\lambda }Z\left( \frac{n\tau }{\lambda ^{2}}+\frac{\kappa }{\lambda }\right) 
\end{equation}
or in other words 
\[
Z^{(n)}(\mathcal{S})=T_{n}Z(\tau )\]
where the \( T_{n} \)-s denote the Hecke-operators well-known from number theory
\cite{Apostol}. So in this case we recover the result of \cite{DMVV}
\begin{equation}
\label{dmvv}
\sum _{n=0}^{\infty }p^{n}Z_{n}(\tau )=\exp \left( Y(p,\tau )\right) 
\end{equation}
with 
\begin{equation}
\label{gcpot1}
Y(p,\tau )=\sum _{n=1}^{\infty }p^{n}T_{n}Z(\tau )
\end{equation}

This last result suggests to interpret the sum in Eq.(\ref{heck}) as a generalization
of the usual Hecke-operators to the higher genus case. But one must take this
identification with a grain of salt, for these operators relate quantities corresponding
to different genera, as an \( n \)-sheeted covering of a genus \( g \) surface
has genus \( n(g-1)+1 \) by the Riemann-Hurwitz formula \cite{Farkas}. 

Along similar lines, one may compute the generating function 
\[
\sum _{n}p^{n}K_{n}(t)\]
of the Klein-bottle amplitudes \( K_{n}(t) \) of the permutation orbifolds
\( \mathcal{C}\wr S_{n} \), which are needed when considering unoriented strings.
In this case we take for \emph{\( \mathcal{S} \)} a Klein-bottle whose partition
function in \( \mathcal{C} \) is \( K(t) \) - the meaning of the positive
parameter \( t \) is that the oriented Schottky cover of the Klein-bottle \( \mathcal{S} \)
is a torus with modular parameter \( \frac{1}{it} \). To apply Eq.(\ref{partgen}),
one has to classify the finite index subgroups of \( G_{\mathcal{S}} \). A
finite index subgroup of \( G_{\mathcal{S}}=\left\langle x,y\, |\, xy=yx^{-1}\right\rangle  \)
is either isomorphic to \( G_{\mathcal{S}} \) itself, in which case the corresponding
covering is a Klein-bottle, or to \( \mathbb {Z}\oplus \mathbb {Z} \), when
the corresponding covering is a torus. In either case, the subgroup can be generated
by \( x^{\lambda } \) and \( x^{\kappa }y^{\mu } \), where one may take \( \lambda ,\mu  \)
to be positive integers and \( 0\leq \kappa <\lambda  \). The index of the
subgroup is \( \lambda \mu  \), and the corresponding covering is a Klein-bottle
(with parameter \( \frac{\lambda t}{\mu } \)) or a torus (with parameter \( \frac{\mu }{2\lambda it}+\frac{\kappa }{\lambda } \)),
according to whether \( \mu  \) is odd or even. The above analysis leads to
the result

\begin{equation}
\label{kleinpart}
\sum _{n=0}^{\infty }p^{n}K_{n}(t)=\exp \left( \mathcal{K}(p,t)+\frac{1}{2}Y\left( p^{2},\frac{1}{it}\right) \right) 
\end{equation}
 where \( Y(\tau ,p) \) denotes the sum in Eq.(\ref{gcpot1}), while 
\begin{equation}
\label{kleinpot}
\mathcal{K}(p,t)=\sum _{n=1}^{\infty }p^{n}\sum _{d|n}\frac{1-(-1)^{d}}{2d}K\left( \frac{nt}{d^{2}}\right) 
\end{equation}

This is not the end of the story, for as was pointed out by Dijkgraaf in \cite{D},
it is possible to introduce a non-trivial discrete torsion \cite{Vafa},\cite{VW},\cite{Douglas}
in the above models, because \( H^{2}(S_{n})=\mathbb {Z}_{2} \) for \( n\geq 4 \).
In our case the presence of discrete torsion amounts to modifying Eq.(\ref{partfun1})
by the introduction of suitable phases 
\begin{equation}
\label{dt}
Z^{\varepsilon }_{n}(\mathcal{S})=\frac{1}{n!}\sum _{\phi :G_{\mathcal{S}}\rightarrow S_{n}}\varepsilon (\phi )\prod _{\xi \in \mathcal{O}(\phi )}\mathcal{Z}(G_{\mathcal{S},\xi })
\end{equation}
 The discrete torsion coefficients \( \varepsilon (\phi ) \) appearing in the
above formula are determined by a 2-cocycle \( \vartheta \in H^{2}(S_{n}) \)
via the following recipe \cite{DW},\cite{D} : each homomorphism \( \phi  \)
determines a pull-back cocycle \( \phi ^{*}\vartheta \in H^{2}(G_{\mathcal{S}}) \).
As the classifying space \( BG_{\mathcal{S}} \) may be identified with the
surface \( \mathcal{S} \), the pull-back cocycle \( \phi ^{*}\vartheta  \)
corresponds to a closed 2-form on \( \mathcal{S} \), and pairing this 2-form
with the fundamental cycle gives the discrete torsion coefficient \( \varepsilon (\phi ) \).
If \( \mathcal{S} \) is a closed orientable surface of genus \( g \), then
a homomorphism \( \phi :G_{\mathcal{S}}\rightarrow S_{n} \) is determined by
the images of the canonical generators of \( G_{\mathcal{S}} \), i.e. \( 2g \)
permutations \( a_{1},b_{1},\ldots ,a_{g},b_{g}\in S_{n} \) that satisfy \( \prod _{j}\left[ a_{j},b_{j}\right] =1 \),
and that represent the monodromies around the cycles of a canonical homology
basis. The corresponding discrete torsion coefficient is given by 
\begin{equation}
\label{dto}
\varepsilon \left( \begin{array}{ccc}
a_{1} & \ldots  & a_{g}\\
b_{1} & \ldots  & b_{g}
\end{array}\right) =\prod _{j=1}^{g}\frac{\vartheta (a_{j},b_{j})\vartheta (a_{j},A_{j-1})}{\vartheta (b_{j},a_{j}^{b_{j}})\vartheta (a_{j},[a_{j},b_{j}])}
\end{equation}
 where \( A_{j}=\prod _{k<j}[a_{j},b_{j}] \) and \( [a,b]=a^{-1}a^{b}=a^{-1}b^{-1}ab \)
is the commutator of \( a \) and \( b \). In particular, for the torus (\( g=1 \))
we get \cite{DW} 
\begin{equation}
\label{dttorus}
\varepsilon _{T}(a,b)=\frac{\vartheta (a,b)}{\vartheta (b,a)}\, \, \, \, \mathrm{for}\, \, ab=ba
\end{equation}
 where \( a \) and \( b \) denote the monodromies around the canonical homology
cycles. The above results follow from an analysis of the irreducible representations
of the second cohomology group \( H^{2}\left( G_{\mathcal{S}}\right)  \). 

In case of unorientable surfaces there is a subtlety, because one should pair
the fundamental cycle not with a 2-form, but with a 2-density, i.e. an element
of the de Rham complex twisted by the orientation bundle \cite{Bott} - in the
orientable case densities are the same as differential forms. This means that
the pull-back cocycle \( \phi ^{*}\vartheta  \) should be a twisted cocycle
of \( \pi _{1}\mathcal{S} \) and not an ordinary one, and this can happen only
if the values of \( \vartheta  \) are restricted to \( \pm 1 \). For a genus
\( g \) non-orientable surface \( \mathcal{S} \), a homomorphism \( \phi :G_{\mathcal{S}}\rightarrow S_{n} \)
is determined by prescribing \( g+1 \) permutations \( v_{0},\ldots ,v_{g}\in S_{n} \)
satisfying \( \prod _{j}v_{j}^{2}=1 \), and the corresponding discrete torsion
coefficient is given by 
\begin{equation}
\label{dtno}
\varepsilon (v_{0},\ldots ,v_{g})=\prod _{j=0}^{g}\vartheta (v_{j}^{2},V_{j})\vartheta (v_{j},v_{j})
\end{equation}
 where \( V_{j}=\prod _{k>j}v_{k}^{2} \). In particular, for the Klein-bottle
we get 
\begin{equation}
\label{dtklein}
\varepsilon _{K}(x,y)=\frac{\vartheta (x,y)\vartheta (x,x^{-1})}{\vartheta (y,x^{-1})}\, \, \, \mathrm{for}\, \, \, xy=yx^{-1}
\end{equation}
 where we wrote the formula in terms of \( x=v_{0}v_{1} \) and \( y=v_{1}^{-1} \)
- this turns out to be more convenient later.

After these generalities, let's turn to the case at hand, i.e. the discrete
torsion corresponding to the non-trivial cocycle \( \vartheta \in H^{2}(S_{n}) \)
for \( n>3 \). As we show in the Appendix, there is a simple closed formula
in this case for the discrete torsion coefficients. To write it down, let's
introduce the quantities 
\begin{equation}
\label{xpar}
\left| x\right| =\sum _{\xi \in \mathcal{O}(x)}\left( \left| \xi \right| -1\right) 
\end{equation}
 and 
\begin{equation}
\label{xypar}
\left| x,y\right| =\sum _{\xi \in \mathcal{O}(x,y)}\left( \left| \xi \right| -1\right) 
\end{equation}
 for arbitrary permutations \( x \) and \( y \), where as usual, we denote
by \( \mathcal{O}(x,y) \) the set of orbits of the subgroup generated by \( x \)
and \( y \), and \( \left| \xi \right|  \) denotes the length of the orbit
\( \xi  \). Then the discrete torsion coefficient for the torus reads 
\begin{equation}
\label{dtt1}
\varepsilon _{T}(x,y)=\left( -1\right) ^{\left( \left| x\right| -1\right) \left( \left| y\right| -1\right) +\left| x,y\right| -1}
\end{equation}
for a pair of commuting permutations \( xy=yx \). For the Klein-bottle coefficients
we have 
\begin{equation}
\label{dtk1}
\varepsilon _{K}(x,y)=(-1)^{(\left| x\right| -1)(\left| y\right| -1)+\frac{\left| x\right| (\left| x\right| +1)}{2}+\left| x,y\right| -1}
\end{equation}
in case \( xy=yx^{-1} \). While looking very similar, these quantities are
in fact quite different, e.g. they do not coincide on the intersection of their
domains of definition.

There is an alternate form for the discrete torsion coefficients that is more
suitable for computations, namely 
\begin{equation}
\label{dtt2}
\varepsilon _{T}(x,y)=(-1)^{\left| x,y\right| }\frac{(-1)^{\left| x\right| }+(-1)^{\left| y\right| }+(-1)^{\left| x\right| +\left| y\right| }-1}{2}
\end{equation}
which follows from the identity 
\begin{equation}
\label{abident}
2(-1)^{ab}+(-1)^{a+b}=(-1)^{a}+(-1)^{b}+1
\end{equation}
valid for integer \( a \) and \( b \). In other words, we have 
\begin{equation}
\label{dtt3}
\varepsilon _{T}(x,y)=\frac{(-1)^{\left| x,y\right| }}{4}\sum _{\alpha ,\beta \in \left\{ \pm 1\right\} }(1-\alpha -\beta -\alpha \beta )\alpha ^{\left| x\right| }\beta ^{\left| y\right| }
\end{equation}
for \( xy=yx \). A similar expression holds for the Klein-bottle coefficients

\begin{equation}
\label{dtk3}
\varepsilon _{K}(x,y)=\frac{(-1)^{\left| x,y\right| }}{4}\sum _{\alpha \in \left\{ \pm i\right\} }\sum _{\beta \in \left\{ \pm 1\right\} }(1-\alpha -\beta -\alpha \beta )\alpha ^{\left| x\right| }\beta ^{\left| y\right| }
\end{equation}
for \( xy=yx^{-1} \). Note that in this last formula the summation variable
\( \alpha  \) ranges over \( \pm i \), while in the torus case its allowed
values were \( \pm 1 \). The virtue of Eqs.(\ref{dtt3}) and (\ref{dtk3})
is that they involve only quantities that are linear in \( \left| x\right|  \)
and \( \left| y\right|  \), unlike the expressions in Eqs.(\ref{dtt1},\ref{dtk1}),
and this makes possible the application of the identity Eq.(\ref{comb}) in
the relevant computations.

Armed with the above, we can now compute the generating functions in the torus
and the Klein-bottle cases. In contrast to the case with trivial discrete torsion,
we no longer get an exponential, rather a combination of four exponential expressions.
The result for the torus reads 
\begin{equation}
\label{dtpart1}
\sum _{n=0}^{\infty }p^{n}Z_{n}^{\varepsilon }(\tau )=\frac{1}{4}\sum _{\alpha ,\beta \in \left\{ \pm 1\right\} }(1-\alpha -\beta -\alpha \beta )\exp \left\{ -Y_{\alpha \beta }(-p,\tau )\right\} 
\end{equation}
where for \( \alpha ,\beta \in \left\{ \pm 1\right\}  \)

\begin{equation}
\label{Yab}
Y_{\alpha \beta }(p,\tau )=\sum ^{\infty }_{n=1}\frac{(\alpha \beta p)^{n}}{n}\sum _{d|n}\sum _{0\leq k<d}\alpha ^{\frac{n}{d}}\beta ^{dk+d+k}Z\left( \frac{n\tau +kd}{d^{2}}\right) 
\end{equation}
Note that a similar result appears in \cite{D}, although in product form, for
the elliptic genus. We also observe that 
\[
Y_{++}(p,\tau )=Y(p,\tau )\]
making contact with the case without discrete torsion. 

The corresponding result for the Klein-bottle reads
\begin{equation}
\label{dtkpart}
\sum _{n=0}^{\infty }p^{n}K_{n}^{\varepsilon }(t)=\frac{1}{4}\sum _{\begin{array}{c}
\alpha \in \left\{ \pm i\right\} \\
\beta \in \left\{ \pm 1\right\} 
\end{array}}(1+\alpha -\beta +\alpha \beta )\exp \left\{ \mathcal{K}_{\alpha \beta }-\frac{1}{2}Y_{-\beta }\left( \beta p^{2},\frac{1}{it}\right) \right\} 
\end{equation}
 where

\begin{equation}
\label{Kab}
\mathcal{K}_{\alpha \beta }(p,t)=i\alpha \sum _{n=1}^{\infty }(\alpha \beta p)^{n}\sum _{d|n}\frac{\left( i^{d}-(-i)^{d}\right) }{4d}\left( \beta ^{\frac{n+d}{2d}}+\beta ^{-\frac{n+d}{2d}}\right) K\left( \frac{nt}{d^{2}}\right) 
\end{equation}

We see that the inclusion of discrete torsion leads to important changes, especially
for the Klein-bottle amplitudes. We omit the derivation of the above expressions,
as it it a rather straightforward but lengthy application of Eq.(\ref{comb})
and some simple arithmetical identities. In principle, it should be possible
to determine similar closed expressions for arbitrary surfaces, but this would
involve the computation of the corresponding discrete torsion coefficients for
higher genera, which does not seem a trivial matter. 

This concludes our survey of \( S_{n} \) permutation orbifolds. We have seen
how one may compute arbitrary partition functions for symmetric products by
suitable use of the combinatorial identity Eq.(\ref{comb}). The analysis goes
through to other quantities of interest as well, showing deep connections with
analytic number theory. As to the physical interpretation of the results, this
is a worthy task well beyond the scope of the present note.

\subsection*{Appendix }

This appendix is devoted to the proof of Eqs.(\ref{dtt1},\ref{dtk1}) for the
discrete torsion coefficients. To this end, let's introduce for \( x,y\in S_{n} \)
the quantity 
\[
\hat{\varepsilon }(x,y)=\left\{ \begin{array}{cc}
\frac{\vartheta (x,y)}{\vartheta (y,x)} & \mathrm{if}\, \, xy=yx\\
\frac{\vartheta (x,y)\vartheta (x,x^{-1})}{\vartheta (y,x^{-1})}(-1)^{\frac{\left| x\right| (\left| x\right| +1)}{2}} & \mathrm{if}\, \, xy=yx^{-1}
\end{array}\right. \]
where \( \vartheta  \) denotes the non-trivial 2-cocycle of \( S_{n} \). The
above quantity is well defined - this needs to be checked for \( x^{2}=1 \),
because in that case \( xy=yx \) is equivalent to \( xy=yx^{-1} \) , so the
two expressions for \( \hat{\varepsilon }(x,y) \) should coincide - because
\( \vartheta (x,x)=(-1)^{\frac{\left| x\right| (\left| x\right| +1)}{2}} \)
for any \( x\in S_{n} \) of order \( 2 \), and satisfies the following conditions
as a consequence of the cocycle relation satisfied by \( \vartheta  \) :

\begin{enumerate}
\item \( \hat{\varepsilon }(x,x)=1 \) 
\item \( \hat{\varepsilon }(x,y)\hat{\varepsilon }(y,x)=1 \) for \( xy=yx \)
\item \( \hat{\varepsilon }(x^{z},y^{z})=\hat{\varepsilon }(x,y) \) 
\item \( \hat{\varepsilon }(x,yz)=\hat{\varepsilon }(x,y)\hat{\varepsilon }(x^{y},z) \)
for \( y,z\in \hat{C}(x)=\left\{ y\, |\, xy=yx^{\pm 1}\right\}  \)
\end{enumerate}
Note that \( \hat{C}(x) \) is a subgroup of \( S_{n} \). Our claim is that
\( \hat{\varepsilon }(x,y) \) equals the quantity 
\[
\varepsilon (x,y):=(-1)^{\left( \left| x\right| -1\right) \left( \left| y\right| -1\right) +\left| x,y\right| -1}\]
for any \( y\in \hat{C}(x) \). This follows if we can show that \( \varepsilon (x,y) \)
satisfies the above conditions for \( \hat{\varepsilon } \) and is non-trivial,
because \( H^{2}(S_{n})=\mathbb {Z}_{2} \) implies that there could be at most
one such quantity.

First, \( \left| x,x\right| =\left| x\right|  \)implies that
\[
\varepsilon (x,x)=(-1)^{(\left| x\right| -1)\left| x\right| }=1\]
because the exponent is always even. On the other hand, the obvious relation
\( \left| y,x\right| =\left| x,y\right|  \) implies that 
\begin{equation}
\label{ators}
\varepsilon (y,x)=\varepsilon (x,y)^{-1}
\end{equation}
 because the values of \( \varepsilon  \) are \( \pm 1 \). That \( \varepsilon (x^{z},y^{z})=\varepsilon (x,y) \)
holds follows from its definition. It remains to show that 
\begin{equation}
\label{ators2}
\varepsilon (x,yz)=\varepsilon (x,y)\varepsilon (x^{y},z)
\end{equation}
 for \( y,z\in \hat{C}(x) \). First, let's note that \( \sigma (x)=(-1)^{\left| x\right| } \)
is the sign of the permutation \( x \), and it is well known that 
\begin{equation}
\label{signum}
\sigma (xy)=\sigma (x)\sigma (y)
\end{equation}
 for any \( x,y\in S_{n} \). But we may rewrite \( \varepsilon  \) in the
form 
\[
\varepsilon (x,y)=\sigma (y)^{\left| x\right| -1}(-1)^{\left| x\right| -\left| x,y\right| }\]

It is straightforward to show that the cyclic subgroup \( \left\langle x\right\rangle  \)
generated by \( x \) is normal in \( \hat{C}(x) \), consequently for any orbit
\( \zeta  \) of \( \hat{C}(x) \), the orbits of \( x \) on \( \zeta  \)
form a block system. In other words, for any orbit \( \zeta \in \mathcal{O}\left( \hat{C}(x)\right)  \),
all \( x \) orbits \( \xi \in \mathcal{O}_{\zeta }(x) \) contained in \( \zeta  \)
have the same length, and there is a homomorphism 
\[
\pi _{\zeta }:\hat{C}(x)\rightarrow Sym\left( \mathcal{O}_{\zeta }(x)\right) \]
 But for \( y\in \hat{C}(x) \) we have 
\begin{equation}
\label{indact}
\left| x\right| -\left| x,y\right| =\sum _{\zeta \in \mathcal{O}(\hat{C}(x))}\left| \pi _{\zeta }(y)\right| 
\end{equation}
 which implies 
\[
\varepsilon (x,y)=\sigma (y)^{\left| x\right| -1}\prod _{\zeta \in \mathcal{O}(\hat{C}(x))}\sigma \left( \pi _{\zeta }(y)\right) \]
 The homomorphism property of \( \pi _{\zeta } \) and Eq.(\ref{signum}) then
imply Eq.(\ref{ators2}).

To complete the proof of \( \hat{\varepsilon }(x,y)=\varepsilon (x,y) \) we
have to show that \( \varepsilon  \) is non-trivial for \( n\geq 4 \), but
this is obvious since \( \varepsilon (x,y)=-1 \) for any two commuting transpositions
\( x,y \).\\
\\
\emph{Work supported by grant OTKA T32453.}

\end{document}